# Bi$_2$Se$_3$ thin films heteroepitaxially grown on *α*-RuCl$_3$

Joon Young Park[1,2], Janghyun Jo[3,*], Jennifer A. Sears[4,†], Young-June Kim[4], Miyoung Kim[3], Philip

Kim[2,‡], and Gyu-Chul Yi[1,§]

[1] *Department of Physics and Astronomy, Institute of Applied Physics, and Research Institute of Advanced Materials, Seoul National University, Seoul 08826, Republic of Korea*

[2] *Department of Physics, Harvard University, Cambridge, Massachusetts 01238, USA*

[3] *Department of Materials Science and Engineering and Research Institute of Advanced Materials, Seoul National University, Seoul 08826, Republic of Korea*

[4] *Department of Physics, University of Toronto, Toronto, Ontario, M5S 1A7, Canada*

[*] Present address: Ernst Ruska-Centre for Microscopy and Spectroscopy with Electrons and Peter Grünberg Institute, Forschungszentrum Jülich, 52425 Jülich, Germany
[†] Present address: Deutsches Elektronen-Synchrotron DESY, 22607 Hamburg, Germany
[‡] Corresponding author: pkim@physics.harvard.edu
[§] Corresponding author: gcyi@snu.ac.kr

Combining various two-dimensional materials into novel van der Waals (vdW) heterostructures has been shown to lead to new emergent quantum systems. A novel heterostructure composed of a vdW topological insulator (TI) such as Bi$_2$Se$_3$ with a quantum spin liquid (QSL) such as *α*-RuCl$_3$ is of great interest for the potential for the chiral Dirac electrons in the TI surface states to interact strongly with the fractionalized fermionic spin excitations in the QSL. We report the heteroepitaxial growth of Bi$_2$Se$_3$ thin films on *α*-RuCl$_3$ as well as the characterization of their structural and electrical properties. Bi$_2$Se$_3$ thin films with an atomically smooth and uniform surface are grown by molecular beam epitaxy. The heterostructure exhibits a preferential epitaxial relationship corresponding to (5 × 5)–Bi$_2$Se$_3$/($2\sqrt{3}$ × $2\sqrt{3}$) *R*30°–*α*-RuCl$_3$ commensurate supercells with a periodicity of 1.2 nm. The formation of the superlattice despite a lattice mismatch as large as 60% is attributed to the van der Waals heteroepitaxy. Magnetotransport measurements as a function of temperature show $Bi_2Se_3$ films grown on $\alpha$-$RuCl_3$ are heavily $n$-doped, $n_e \sim 10^{14}$ cm$^{-2}$, with mobility $\mu \sim 450$ cm$^2$ V$^{-1}$ s$^{-1}$ at low temperatures.

# I. INTRODUCTION

Van der Waals (vdW) heterostructures composed of various two-dimensional (2D) layered materials are versatile platforms to create unique materials properties for exploring a wide variety of interesting physical properties and device applications [1–3]. The $(Bi,Sb)_2(Te,Se)_3$ material family are well-known vdW topological insulators (TIs) [4,5]. When these TIs form novel heterointerfaces with other materials in a vdW heterostructure, it has the potential to generate unique topological electronic structures for application in next generation spintronics and quantum computation [6]. The interaction between TIs and other vdW materials such as graphene, hexagonal boron nitride (h-BN), 2D layered ferromagnets and superconductors has been investigated intensively as a source of novel states of matter [7–15]. A new topological material of great interest is quantum spin liquids (QSL). In QSLs, interacting spins create long-range quantum entanglement with fractionalized fermionic excitations without ordinary magnetic order [16,17]. $α$-RuCl$_3$ is a promising candidate for a QSL that has become an intense focus of current research. [18–22]. The layered vdW lattice structure of $α$-RuCl$_3$ makes it possible to combine atomically thin $α$-RuCl$_3$ with other layered materials to form vdW heterostructures [23]. The recent demonstration of $α$-RuCl$_3$/graphene heterostructures suggests strong charge transfer across the $α$-RuCl$_3$/graphene interface [24,25]. In TI/graphene heterostructures, it has been shown that the strong spin-orbit coupling (SOC) of the TI can be transferred to graphene by means of proximity effect, leading to a significant enhancement of SOC in graphene by several orders of magnitude [26–28]. Similarly, the proximity SOC from the TI surface states may alter the spin Hamiltonian of $α$-RuCl$_3$, significantly modifying the dynamics of fractionalized fermionic spin excitations in the QSL.

In this Article, we report on the heteroepitaxial growth of $Bi_2Se_3$ on $\alpha$-$RuCl_3$ layers using molecular beam epitaxy (MBE). The hexagonal in-plane lattice symmetry of $\alpha$-$RuCl_3$ and $Bi_2Se_3$ allows highly crystalline $Bi_2Se_3$ film growth by means of heteroepitaxy [29,30]. We use a transmission electron microscope (TEM) and magnetotransport to characterize the atomic structure and electronic properties of this TI/QSL heterostructures.

## II. EXPERIMENTAL METHODS

Heteroepitaxial growth is one of the most effective approaches to fabricating uniform materials with well-defined structural properties. We employ an ultra-high vacuum (UHV) MBE system to grow high quality $Bi_2Se_3$ thin films heteroepitaxially on $\alpha$-$RuCl_3$ single crystal surfaces. The $\alpha$-$RuCl_3$ layers are prepared on $SiO_2$/Si substrates by mechanical exfoliation of $\alpha$-$RuCl_3$ bulk single crystals *in-situ* in high vacuum. The lateral size of $\alpha$-$RuCl_3$ crystallites typically is in the range of several to tens of micrometers while the thickness is tens of nanometers. Before $Bi_2Se_3$ growth, gentle thermal cleaning is carried out at 350 °C for 15 min in UHV. High purity Bi (99.999%, Alfa Aesar) and Se (99.999+%, Alfa Aesar) are co-evaporated by effusion cells (Effucell). The growth is performed under a beam flux ratio of 1:17 Bi to Se, and the growth rate is ~0.3 quintuple layer (QL)/min. The pressure of the MBE chamber is maintained at $10^{-10}$ Torr during the growth.

We adopt the two-stage growth method that was previously developed on h-BN crystals. This method enables increased nucleation density on the vdW substrate without causing texturing of the film [31]. In the first stage, we deposit the initial 2–3 QLs of $Bi_2Se_3$ at a temperature of 180 °C. The substrate temperature is then slowly elevated at a rate of 5°C/min under Se flux. For the second stage, the ultrathin film is annealed for 30 min at a higher temperature (300–350°C), and then the growth is resumed until the desired thickness is reached. The thickness of $Bi_2Se_3$ films investigated in this work is 20 QLs.

The structural characteristics of $Bi_2Se_3$ grown on $\alpha$-$RuCl_3$ ae investigated by atomic force microscopy (AFM) and plan-view and cross-sectional TEM (JEM-2100F, JEOL). For plan-view TEM sampling, as-grown $Bi_2Se_3/\alpha$-$RuCl_3$ layers are transferred onto TEM grids by poly(methyl methacrylate) (PMMA) spin coating followed by wet chemical etching of the sacrificial $SiO_2$ layer in $SiO_2$/Si substrates. Subsequently, the PMMA support layer is removed by immersion in an acetone bath, and the sample is rinsed with isopropyl alcohol. The cross-sectional TEM specimen is prepared by focused ion beam milling (Helios 650 FIB, FEI) and low-energy focused Ar ion milling (NanoMill 1040, Fischione).

Electrical transport measurements are performed by fabricating the material into a Hall bar or van der Pauw device according to the following steps. First, an array of Cr/Au 5/50 nm microdots are deposited by electron beam (*e*-beam) evaporation through stencil masks to form ohmic contacts. Then we spin coat hydrogen silsesquioxane (HSQ) layer on top of a sacrificial layer of PMMA. E-beam lithography is used to pattern the HSQ layer into the desired device geometry. The exposed PMMA layer is etched by oxygen plasma. The exposed $Bi_2Se_3$ films are etched using Ar plasma. The PMMA and HSQ mask is removed with acetone. Bonding pads to connect to the device are fabricated with *e*-beam lithography followed by *e*-beam evaporation of Cr/Pd/Au 5/20/125 nm. Finally, the sample is capped with a layer of PMMA, and the bonding pads are exposed by another round of *e*-beam lithography. The magnetotransport measurements are carried out in a Physical Property Measurement System from Quantum Design by a low-frequency AC lock-in technique with an excitation current of 500 nA.

## III. RESULTS AND DISCUSSION

The surface morphology of $Bi_2Se_3$ thin films grown on $α$-$RuCl_3$ is characterized by AFM. Figures 1(a)–(c) display AFM topography images of the films for different temperatures of the second stage of growth. They show triangular crystal facets and terrace structures of $Bi_2Se_3$. For a second stage growth temperature of 300°C, the film [Fig. 1(a)] exhibits a surface decorated with faint facets and sub-100 nm terraces. The root-mean-square (RMS) roughness value for the 2 × 2 µm$^2$ area is 0.98 nm. When the second stage temperature is increased to 330°C, we find a smoother surface with the RMS roughness of 0.59 nm [Fig. 1(b)]. The crystallographic facets are also more evident, and the terraces are larger when the temperature of the second stage of growth is increased to 330°C in Fig. 1(b) compared to the lower temperature shown in Fig. 1(a). When we increase the temperature for the second stage of growth even further to 350°C [Fig. 1(c)], the $Bi_2Se_3$ domains are discontinuous and $Bi_2Se_3$ nanoplatelets emerge. The line profile in Fig. 1(c) displays a height step of ~20 QL between isolated $Bi_2Se_3$ nanoplatelets. Based on the observed growth behavior, we chose 350°C as the optimal growth temperature for the second stage, which yields thin films with a uniform and flat surface morphology. The lateral size of the atomically smooth terrace is several hundreds of nanometers [see the line profile in Fig. 1(b)]. We remark, however, that the observed terrace size is smaller by a factor of 2–4 than that of $Bi_2Se_3$ grown on h-BN at similar conditions [31]. The reduced single crystal domain size is presumably due to differences in lattice constant mismatches and interactions at the vdW interface, as described below.

We investigated the heteroepitaxial relationship of $Bi_2Se_3$ and $α$-$RuCl_3$ using plan-view selected-area electron diffraction (SAED). Figure 2(a) is a typical SAED image obtained from a circular illumination area with a diameter of 4.1 µm. The SAED image exhibits a set of hexagonally symmetric bright spots and another set of hexagonal, broadened spots accompanied by faint arc backgrounds. Based on the lattice parameter analysis, the former corresponds to the single crystalline $\alpha$-RuCl$_3$ substrate and the latter corresponds to the aligned Bi$_2$Se$_3$ crystallites grown on top of it. As for the statistical distribution of the in-plane alignment of Bi$_2$Se$_3$ domains, the standard deviation of the SAED intensity profile of Bi$_2$Se$_3$ (11$\bar{2}$0) is estimated to be narrower than $\pm 1.8°$ [Fig. 2(b)]. The preferential orientation of Bi$_2$Se$_3$ (11$\bar{2}$0) is aligned parallel to $\alpha$-RuCl$_3$ (10$\bar{1}$0), and is marked by the red arrow and the green circle in Fig. 2(a), respectively. The distribution of the Bi$_2$Se$_3$(11$\bar{2}$0) peak suggests a deviation from the perfect epitaxial relationship, indicating the weekly bound attraction of the vdW epitaxial Bi$_2$Se$_3$/$\alpha$-RuCl$_3$ heterointerface.

High-resolution TEM (HR-TEM) reveals the microstructural properties of the Bi$_2$Se$_3$/$\alpha$-RuCl$_3$ heterostructure. Plan-view HR-TEM image in Fig. 2(c) shows clear periodic triangular patterns with a periodicity of ~1.2 nm originating from the moiré fringes due to the large lattice mismatch between Bi$_2$Se$_3$ and $\alpha$-RuCl$_3$. These interference patterns have a periodicity corresponding to the double diffraction between Bi$_2$Se$_3$ {11$\bar{2}$0} and $\alpha$-RuCl$_3$ {30$\bar{3}$0} planes — the strongest diffraction peaks of each layer [see Fig. 2(a)] — marked by the blue dotted circle in Fig. 2(a). To distinguish the lattice of each layer from this overlapped image, we reconstruct the individual lattice images through inverse FFT processing, using the two strongest sets of diffraction patterns. The resulting image, shown in the lower left (right) inset in Fig. 3(c), reveals a lattice spacing of 0.21 (0.17) nm, agreeing well with the spacing of the {11$\bar{2}$0} ({30$\bar{3}$0}) planes of Bi$_2$Se$_3$ ($\alpha$-RuCl$_3$) which corresponds to $a_{\text{Bi}_2\text{Se}_3}/2$ ($\sqrt{3}a_{\alpha\text{-RuCl}_3}/6$) [29,30].

In order to probe the interfacial quality, we further examine the out-of-plane lattice structures of the Bi$_2$Se$_3$/$\alpha$-RuCl$_3$ using cross-sectional HR-TEM, as shown in Fig. 2(d). The heterointerface is unambiguously distinguished as marked by a yellow dashed line. The vertically stacked layered structures of Bi$_2$Se$_3$ and $\alpha$-RuCl$_3$ are also clearly observed. The lattice spacings between adjacent planes in $Bi_2Se_3$ and $\alpha$-$RuCl_3$, indicated by parallel white and black dashed lines, are measured to be 0.95 and 0.57 nm, which agree well with reported *d*-spacings of $Bi_2Se_3$ (0003) [29] and $\alpha$-$RuCl_3$ (0003) [30], respectively. The image also shows that the $Bi_2Se_3$ film is highly crystalline down to the first quintuple layer, and displays an atomically sharp interface with the underlying $\alpha$-$RuCl_3$ layers. No extended crystal defects such as threading dislocations are observed at the interface. We also note, however, that $\alpha$-$RuCl_3$ is extremely fragile against high-energy processes such as ion milling for cross-sectional sampling. We find that $\alpha$-$RuCl_3$ is easily amorphized by the electron beam irradiation during HR imaging, making further extended and detailed cross-sectional analysis challenging.

Based on our TEM observations, we simulate a ball-and-stick lattice model to illustrate the atomic configuration of the epitaxial $Bi_2Se_3$/$\alpha$-$RuCl_3$ heterostructure [32]. As both crystals are composed of planar layers vertically stacked through weak vdW interactions [29,30], in-plane structures of the single layer components that form the interface are displayed [Figs. 3(a) and (b)]. While the lattice parameter of $Bi_2Se_3$ is significantly smaller than that of $\alpha$-$RuCl_3$ by ~60%, the in-plane lattice periodicities of $Bi_2Se_3$ and $\alpha$-$RuCl_3$ along $[1\bar{2}10]_{Bi_2Se_3} \parallel [1\bar{1}00]_{\alpha\text{-}RuCl_3}$ — $a_{Bi_2Se_3}$ and $\sqrt{3}a_{\alpha\text{-}RuCl_3}$, respectively — match almost perfectly to a small integer ratio of 2:5 with a misfit smaller than 0.1%. Figure 3(c) shows a ball and stick model of the superlattice where $Bi_2Se_3$ and $\alpha$-$RuCl_3$ are aligned along the $[1\bar{2}10]_{Bi_2Se_3} \parallel [1\bar{1}00]_{\alpha\text{-}RuCl_3}$ direction. This model is in good agreement with the plan-view TEM observations, as marked by the orange dotted arrow in the SAED image [Fig. 2(a)]. A faint arc pattern originating from $Bi_2Se_3$ ($22\bar{4}0$) overlaps with the $\alpha$-$RuCl_3$ ($50\bar{5}0$) diffraction spot. Therefore, these observations strongly suggest the formation of a (5 × 5)–$Bi_2Se_3$/($2\sqrt{3} \times 2\sqrt{3}$) $R30°$–$\alpha$-$RuCl_3$ commensurate supercell, which is displayed as red dotted lines in Fig. 3(c). The 30° twist alignment of $Bi_2Se_3$ and $α$-$RuCl_3$ also produces triangular moiré patterns with a periodicity of 1.2 nm [blue dashed triangles in Fig. 3(c)]. Due to the atomic scale relaxation across the vdW interface, a supercell with the moiré periodicity can occur at the vdW interface [33]. Figure 2(c) clearly shows triangular moiré patterns with a periodicity of 1.2 nm in the in-plane TEM image, further supporting $[1\bar{2}10]_{Bi_2Se_3} \parallel [1\bar{1}00]_{α\text{-}RuCl_3}$ vdW epitaxy between $Bi_2Se_3$ and $α$-$RuCl_3$.

Our atomic structural analysis above indicates that the two lattices spontaneously form a large-scale periodicity, whose area is an order of magnitude larger than that of unit cells, despite the substantial lattice mismatch between unit cells. A similar interfacial alignment is reported in $Bi_2Se_3$/h-BN [31] and ZnO/h-BN [34] heteroepitaxial systems where two dissimilar hexagonal lattices with large lattice mismatch are bound by vdW-type interactions. However, the 30°-rotated alignment of $[1\bar{2}10]_{Bi_2Se_3} \parallel [1\bar{1}00]_{α\text{-}RuCl_3}$ is distinctly different from the parallel (0°) alignments of $[1\bar{1}00]_{Bi_2Se_3 (ZnO)} \parallel [1\bar{1}00]_{h\text{-}BN}$ [31,34]. Since $α$-$RuCl_3$, $Ru^{3+}$ atoms in the edge-sharing $RuCl_6$ octahedra form a honeycomb network analogous to the h-BN lattice [Fig. 3(b)] [30,35], the difference in the epitaxial relationship between $Bi_2Se_3$/$α$-$RuCl_3$ and $Bi_2Se_3$/h-BN can be correlated with the commensurate (incommensurate) nature of the supercell for the former (latter) case. The moiré structure formed at the $Bi_2Se_3$/$α$-$RuCl_3$ interface can lead to the emergent quantum states due to the periodic potential modulation provided by the moiré superlattice [36].

The electrical properties of the $Bi_2Se_3$/$α$-$RuCl_3$ heterostructure are investigated by multi-terminal magnetotransport measurements. The longitudinal and Hall resistances, $R_{xx}$ and $R_{xy}$, are measured in Hall-bar devices fabricated from the heterostructure [inset of Fig.4(a)]. Figure 4(a) shows magnetoresistance ratio (MR; $R_{xx}(B)/R_{xx}(B = 0$ T$)$) under magnetic field $B$. When the field is in the out-of-plane direction, MR exhibits a weak positive magnetoresistance (~25%) up to 14

T at low temperatures. The origin of this positive MR potentially can be assigned to the weak antilocalization in TIs [37–41]. The in-plane field MR, however, does not exhibit significant variation up to 14 T. The absence of an in-plane field MR suggests that carrier transport in this heterostructure is two-dimensional, presumably dominant at the $Bi_2Se_3/\alpha$-$RuCl_3$ interface. We also observe no appreciable change of the carrier transport behaviors in the low temperature and high in-plane magnetic field (> 7 T) regime where the transition of $\alpha$-$RuCl_3$ into the QSL state might take place [42–45].

The Hall resistance $R_{xy}(B)$ measured under a perpendicular magnetic field at various temperatures exhibits linear field dependence with a negative slope (data not shown), indicating the dominance of a single $n$-type conduction channel in our device. Figure 4(b) shows the temperature ($T$) dependence of sheet carrier concentration and mobility $\mu$ calculated from the magnetotransport data. Carrier concentration does not decrease appreciably when the temperature is reduced, indicating full ionization of donors (~$8 \times 10^{13}$ cm$^{-2}$) in the film even at low temperatures [46,47]. $\mu$ increases as $T$ decreases and saturates at ~450 cm$^2$ V$^{-1}$ s$^{-1}$ for temperatures below $T$ ~30 K, suggesting that the dominant scattering is originating from static disorders such as the Se vacancy or grain boundaries in the $Bi_2Se_3$ film [48,49]. In light of the relatively high carrier density [50] and that $\alpha$-$RuCl_3$ alone exhibits insulating temperature-dependent resistance behavior [51,52], we attribute the dominant conduction channel to the bulk states and/or impurity bands in $Bi_2Se_3$. Due to the highly conductive parallel conduction channels formed in these states, transport through the TI/$\alpha$-$RuCl_3$ interfacial states is not readily observed in our experiments. The low mobility and high carrier density of $Bi_2Se_3$ grown on $\alpha$-$RuCl_3$, compared with $Bi_2Se_3$/h-BN, can presumably be due to its smaller terrace sizes and larger deviation in the in-plane alignment of $Bi_2Se_3$ grains [31].

## IV. CONCLUSION

In summary, we have demonstrated the heteroepitaxial growth of $Bi_2Se_3$ thin films on $\alpha$-RuCl$_3$ layers. Uniform films are prepared through optimization of the two-stage MBE growth process. The resulting film exhibits atomically smooth surface morphology with the lateral terrace size exceeding hundreds of nanometers. TEM analysis shows an epitaxial relationship of $[1\bar{2}10](0001)_{Bi_2Se_3} \parallel [1\bar{1}00](0001)_{\alpha\text{-RuCl}_3}$ where van der Waals hetero-epitaxy corresponds to the $(5 \times 5)$–$Bi_2Se_3/(2\sqrt{3} \times 2\sqrt{3})$ $R30°$–$\alpha$-RuCl$_3$ commensurate supercell. Magnetotransport measurement shows dominance of $n$-type carriers, likely originating from impurity states in $Bi_2Se_3$, with low-temperature carrier mobility of 470 cm$^2$ V$^{-1}$ s$^{-1}$. The TI/QSL heteroepitaxy realized in our study can provide a route to building material platforms for novel quantum heterostructures by the MBE enabled vdW interface engineering.

**ACKNOWLEDGMENTS**

The authors would like to thank T. S. Mentzel for help in the preparation of the manuscript. This work was financially supported by the Global Research Laboratory Program (2015K1A1A2033332) through the National Research Foundation of Korea (NRF) funded by the Ministry of Science and ICT (MSIT). PK acknowledges support from the DoD Vannevar Bush Faculty Fellowship N00014-18-1-2877 J. Y. Park is grateful to Global Ph.D Fellowship Program through the NRF funded by the Ministry of Education (2012H1A2A1003288). J. Y. Park and G. -C. Yi acknowledge technical support from Effucell Co. J. Jo and M. Kim acknowledge the Creative Materials Discovery Program through NRF (2017M3D1A1040688) and the NRF grant funded by the MSIT (2017R1A2B3011629). Work at the University of Toronto was supported by Natural Science and Engineering Research Council (NSERC) of Canada through Discovery Grant and CREATE program.

**Figure 1**

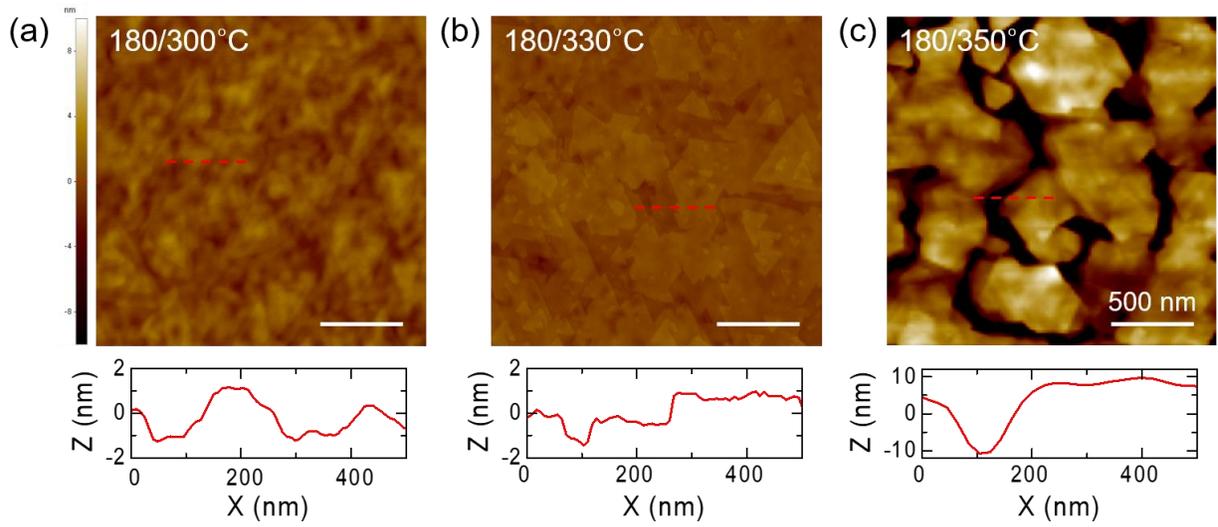

FIG. 1. AFM topography images of 20 QL $Bi_2Se_3$ thin films grown on $\alpha$-RuCl$_3$ at various temperatures for the second stage of growth: (a) 300°C, (b) 330°C, and (c) 350°C. Lower panels show line height profiles taken from the red dashed lines in the upper image. The height profile in (b) exhibits 1 QL steps in $Bi_2Se_3$, corresponding to 1 nm.

**Figure 2**

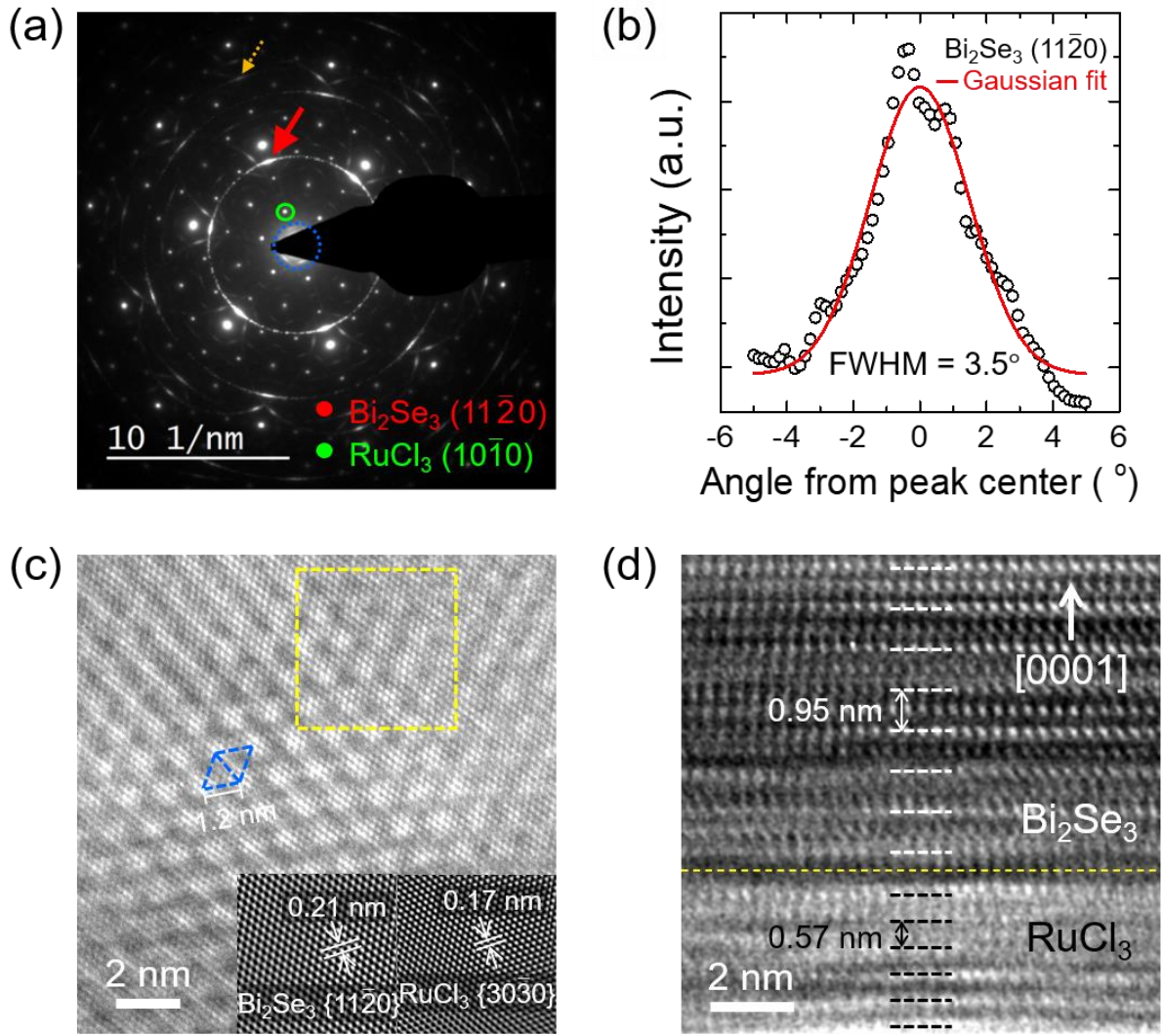

FIG. 2. Microstructural properties of $Bi_2Se_3/\alpha$-$RuCl_3$ investigated by TEM. (a) SAED pattern obtained from a circular illumination area with a diameter of 4.1 μm. (b) Intensity profile of the $Bi_2Se_3$ (11$\bar{2}$0) diffraction and its Gaussian fit exhibiting the full width at half maximum of 3.5°. (c) Plan-view HR-TEM image of $Bi_2Se_3/\alpha$-$RuCl_3$. Moiré fringes, marked by blue dashed triangles, are observed. The lower left (right) inset shows a Fourier-filtered image reconstructed from $Bi_2Se_3$ {11$\bar{2}$0} ($\alpha$-$RuCl_3$ {30$\bar{3}$0}) set of the FFT patterns, corresponding to the area marked by the yellow square in the main figure. (d) Cross-sectional HR-TEM image of the $Bi_2Se_3/\alpha$-$RuCl_3$ heterointerface taken along the $Bi_2Se_3$ [11$\bar{2}$0] direction. An average background subtraction filter was applied in (c) and (d) to remove noise.

**Figure 3**

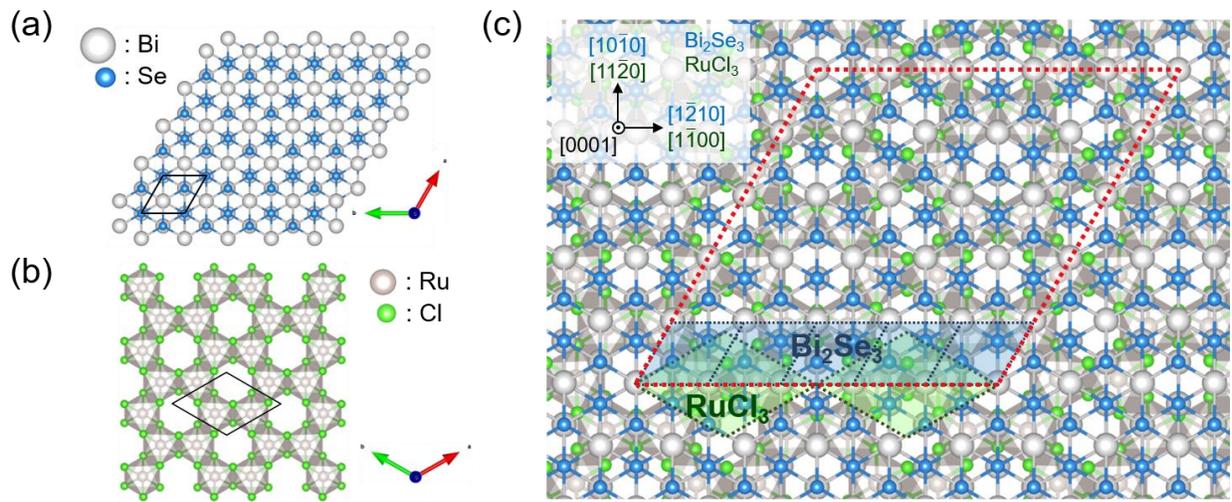

FIG. 3. Ball-and-stick model of (a) $Bi_2Se_3$ and (b) $\alpha$-$RuCl_3$. Primitive unit cells are indicated by solid lines. (c) Epitaxial $Bi_2Se_3$/$\alpha$-$RuCl_3$ heterostructure. The (5 × 5)–$Bi_2Se_3$/($2\sqrt{3} \times 2\sqrt{3}$) $R30°$–$\alpha$-$RuCl_3$ supercell (red dotted line), moiré patterns (blue dashed triangles), and primitive unit cells of $Bi_2Se_3$ (blue shaded area) and $\alpha$-$RuCl_3$ (green shaded area) are indicated.

**Figure 4**

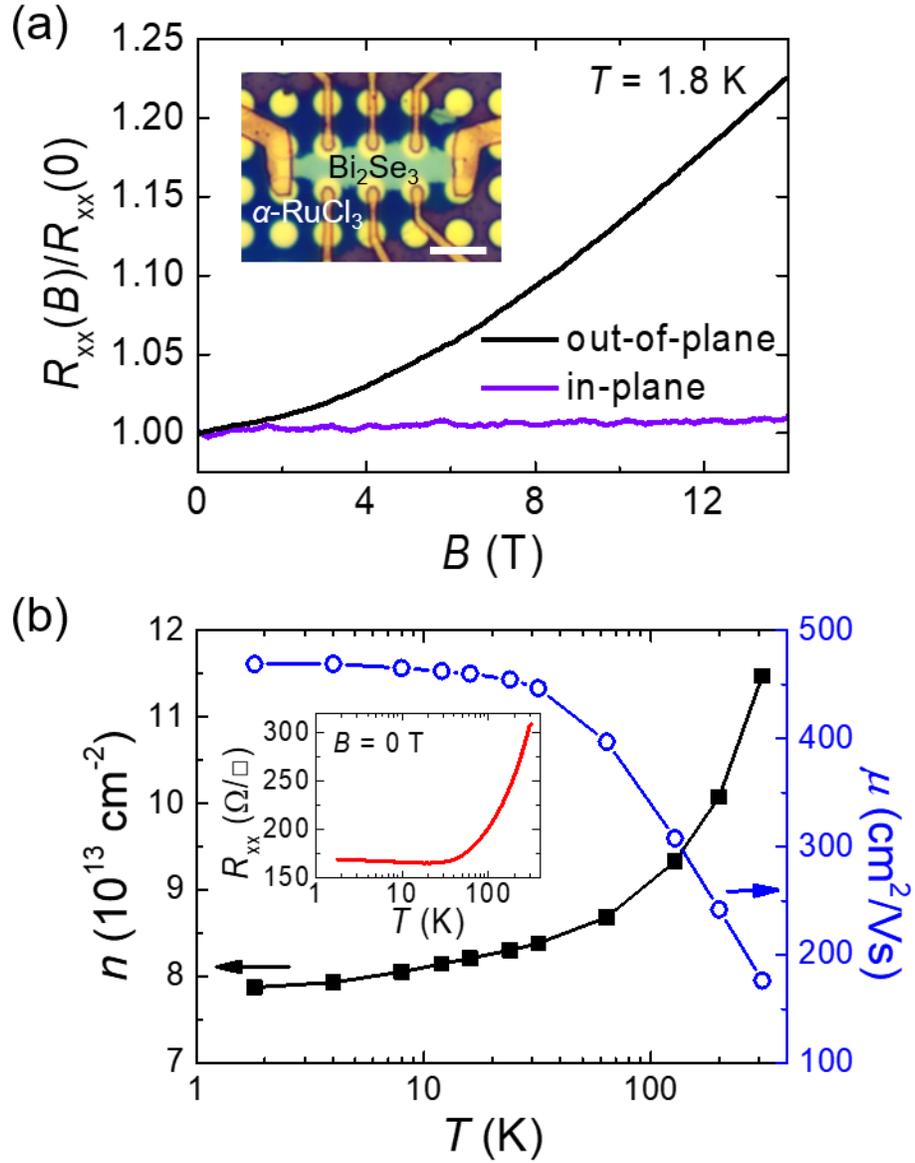

FIG. 4. Electrical transport properties of $Bi_2Se_3/\alpha\text{-RuCl}_3$. (a) Magnetoresistance ratio ($R_{xx}(B)/R_{xx}(0)$) in out-of-plane and in-plane magnetic field $B$ at 1.8 K. Inset: Optical image of a typical Hall bar device fabricated from $Bi_2Se_3/\alpha\text{-RuCl}_3$ heterostructures. Scale bar: 5 μm (b) Temperature dependence of the carrier density ($n$) and mobility ($\mu$) calculated by the Hall measurement in out-of-plane fields. Inset: Temperature dependence of the resistance of $Bi_2Se_3/\alpha\text{-RuCl}_3$ at zero magnetic field.